# Nuclear magnetic resonance spectroscopy of single subnanoliter ova

Marco Grisi[a,1], Beatrice Volpe[b], Roberto Guidetti[c], Nicola Harris[b], Giovanni Boero[a,1]

[a]Microengineering Institute, École Polytechnique Fédérale de Lausanne (EPFL), Lausanne, 1015, Switzerland
[b]Global Health Institute, École Polytechnique Fédérale de Lausanne (EPFL), Lausanne, 1015, Switzerland
[c]Department of Life Sciences, University of Modena and Reggio Emilia, Modena, 41125, Italy

**Nuclear magnetic resonance (NMR) spectroscopy is, in principle, a promising candidate to study the intracellular chemistry of single microscopic living entities. However, due to sensitivity limitations, NMR experiments were reported only on very few and relatively large single cells down to a minimum volume of 10 nl. Here we show NMR spectroscopy of single ova at volume scales (0.1 and 0.5 nl) where life development begins for a broad variety of animals, humans included. We demonstrate that the sensitivity achieved by miniaturized inductive NMR probes (few pmol of $^1$H nuclei in some hours at 7 T) is sufficient to observe chemical heterogeneities among subnanoliter ova of tardigrades. Such sensitivities should allow to non-invasively monitor variations of concentrated intracellular compounds, such as glutathione, in single mammalian zygotes.**

Nuclear magnetic resonance (NMR) is a well-established spectroscopic technique widely employed in physics, chemistry, medicine, and biology. It allows experiments on living matter (1, 2), whose relevance in biology is proven by developments such as *in vivo* protein structure determination (3), metabolic profiling (4), visualization of gene expression (5), and latent phenotype characterization (6). Despite its advantages, NMR suffers from a significantly lower sensitivity with respect to other methods. As a result, experiments are often restricted to large numbers of cells or microorganisms (1, 3, 4, 6). Single cell studies are necessary to fully understand phenomena such as the metabolic heterogeneity within a cell population (7-9). Recently, a number of techniques were applied to intracellular metabolic profiling at single cell scale, all having different limitations and degree of invasivity. For instance, mass spectrometry and fluorescence labeling allow high sensitivities, but require cellular content extraction or labeling with fluorophores (7, 9). Questions concerning invasivity stimulated the coin of the biological equivalent of the so called observer effect, referring to the inability to separate a measurement from its potential influence on an observed cell (9). In this regard, NMR is one of the most promising technique for studies of intracellular compounds in untouched living microorganisms (i.e., with an extremely low level of physical and chemical perturbation) (1, 7).

The first single-cell NMR experiments were performed on *Xenopus laevis* ova (10) which have volumes of approximately 1 μl. Later, it was reported the study of single giant neurons of *Aplysia californica* (11), with volumes of approximately 10 nl. The particularly large volumes of these cells allowed profiling of highly concentrated metabolites and their subcellular localization (12, 13), imaging of *Xenopus laevis* cleavage (14) and neurons structure (15), and study of water diffusion properties within the cytoplasm and nucleus (10, 11, 16).

Here we show NMR experiments on single nematode *Heligmosomoides polygyrus bakeri* (*Hp*) and tardigrade *Richtersius coronifer* (*Rc*) ova. *Rc* and *Hp* ova are just two of the many models present at the subnanoliter scale (Fig. 1A), which include numerous species of microorganisms, echinoderms, and mammals (humans included) (17). *Rc* ova appear as spherical with conical processes on the cuticular surface of the egg shell and have a typical volume of 0.5 nl (Fig. 1B). Their embryonic development is relatively slow, with the eggs hatching in more than 50 days (18). *Hp* ova are ellipsoidal and have a typical volume of about 0.1 nl (Fig. 1C). Fecundated ova of *Hp* develop into a fully embryonated state within 24 hours and within two days stage 1 larvae begin to emerge (19).

NMR spectroscopy of subnanoliter biological samples is both a volume and concentration limited problem, setting severe constraints on the required spin sensitivity. In this work we employ a recently developed single-chip integrated NMR probe (20) where the combination of a low noise transceiver and a multilayer microcoil allows high spin sensitivities in subnanoliter volumes (Fig. 1D). Our work suggest the use of miniaturized high sensitivity NMR probes as a possible approach to study the intracellular chemistry of subnanoliter zygotes.

## Results

Figure 1E describes the assembled probe. A single ovum is first isolated on the surface of an agarose gel contained in a cylindrical polystyrene cup. Later, the cup is positioned on top of the microchip in such a way that the ovum is precisely placed over the microcoil. The cup is fixed to the printed circuit board with candle wax. The gel keeps the sample in close contact with the coil for days without physically damaging it, whereas the wax prevents gel drying. The microchip is wire bonded to a printed circuit board and inserted in the room temperature bore of a 7.05 T (300 MHz) superconducting magnet. The bonding wires are electrically isolated by a silicone glue.

Figure 2A shows a collection of single *Rc* ova $^1$H NMR spectra resulting from 12 hours averaging. The shimming coils system of the magnet at these volume scales is highly ineffective, resulting in measured linewidths of about 50 Hz due to susceptibility mismatches in the sample region. In these measurements, the agarose gel is obtained by dispersing 1.5% agarose in either M9 buffer, $H_2O$, or $D_2O$. Single ova (a), (b), and (c) as indicated in Fig. 2A are measured in $H_2O$ based gels. In these spectra only slight differences in signal amplitudes (Fig. S1A) and relative abundances (Fig. S1B) are visible.

As shown in Fig. 2A, the water signal overlaps with nearby resonance lines. In these conditions, water suppression techniques can be hardly applied without significant spectral artifacts (21).

[1]giovanni.boero@epfl.ch, marco.grisi@epfl.ch

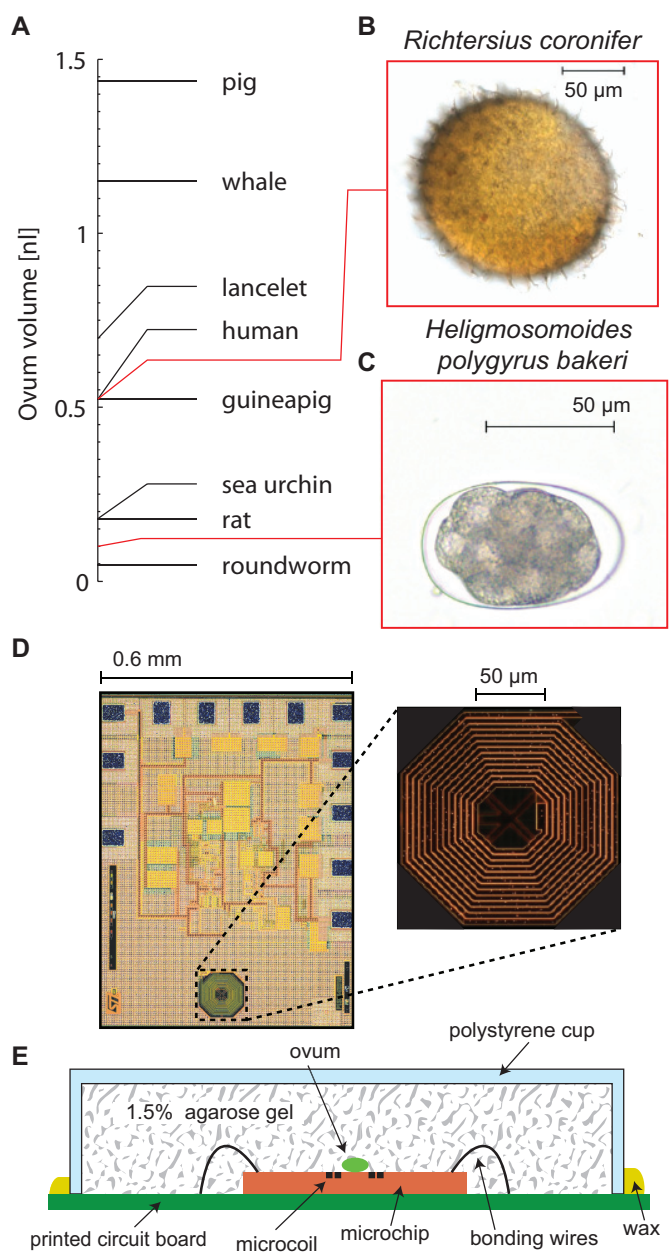

Fig. 1. Samples and setup. (A) Approximate volumes of ova of selected animals. (B-C) Optical images of the ova studied in this work. (D) Photographs of the integrated microchip and microcoil (see details in Ref.(20)). (E) Schematic representation of the single ovum probe in section view.

With the aim of having a significantly smaller background signal, we investigated the use of heavy water ($D_2O$) as an alternative to water. In $D_2O$ based gels, HDO is formed by proton exchange with the OH groups in the agarose molecules. HDO resonates at about 0.03 ppm relative to the $H_2O$ chemical shift (22) and constitutes the only background signal that is visible in our experimental conditions and time scales (Fig. S2). The weaker background signal in $D_2O$ gels (about 50 times smaller than in $H_2O$ gels) allows to better resolve the resonance lines close to water and can be used as internal chemical shift reference.

Figures 2B and 2C shows NMR spectra of single *Rc* ova in $D_2O$ gels. These spectra indicate that the *Rc* ova have a certain degree of heterogeneity. In general, three kinds of heterogeneity among NMR spectra can be present, each one linked to different sample properties. Heterogeneity in linewidths is caused by differences in the spin-spin relaxation time or field inhomogeneity within the sample, heterogeneity in absolute signal amplitudes is due to differences in the amount of a substance within the sample, and heterogeneity in relative signal amplitudes is a consequence of differences in the chemical composition of the sample.

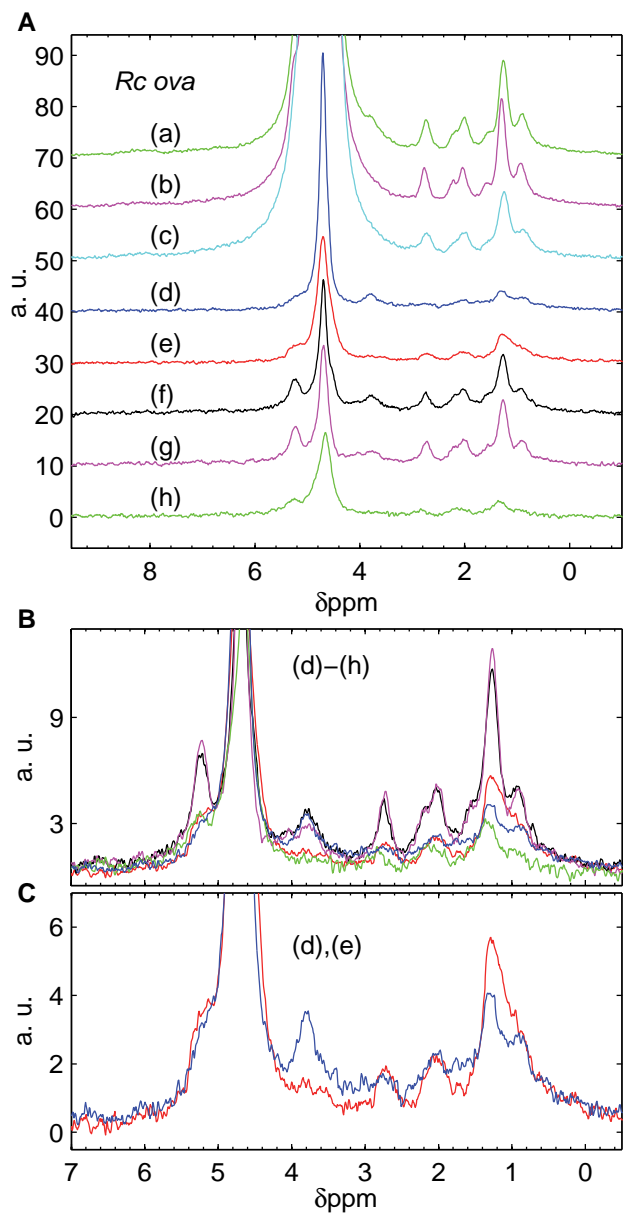

Fig. 2. NMR spectra of single *Richtersius coronifer* (*Rc*) ova. (A) Eight *Rc* single ovum experiments in M9 (a), $H_2O$ (b-c), and $D_2O$ (d-h) based gels. (B) Detailed comparison of *Rc* ova in $D_2O$ based gels. (C) Detailed comparison of two *Rc* ova in $D_2O$ based gels. Colors in 2B and 2C refer to the same ova as in 2A.

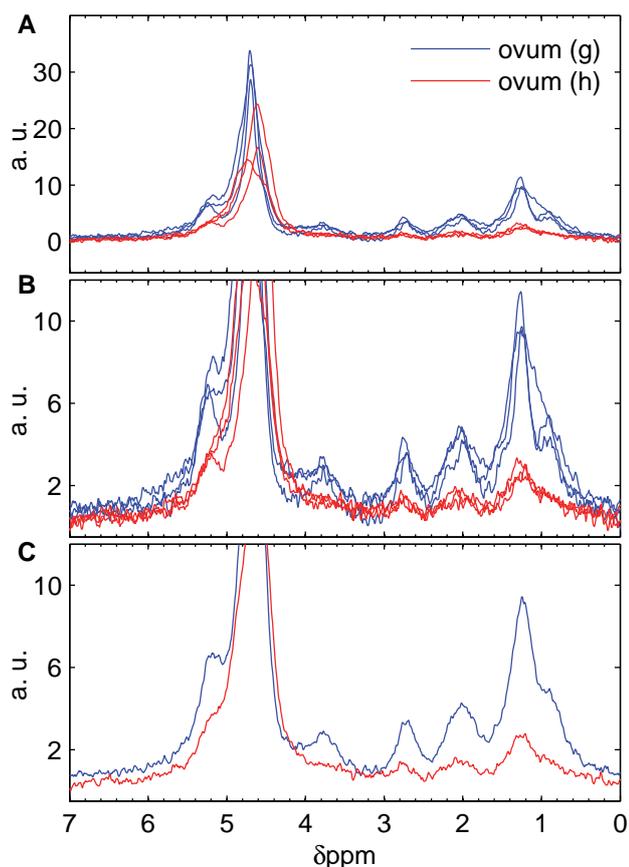

Fig. 3. Comparison between measurements of two *Rc* ova, ovum (g) and ovum (h) as indicated in Fig. 2A. Each ovum was arbitrarily repositioned three times and measured. (A) NMR spectra. Detailed comparison of all spectra (B) and averaged spectra (C).

It is important to note that the volume in which our excitation/detection microcoil has an approximately uniform sensitivity roughly corresponds to a half-sphere of 100 μm diameter (i.e., about 0.25 nl). With their volumes of about 0.5 nl, *Rc* ova exceed this region. In order to investigate if the origin of the observed heterogeneity is due to a non-uniform chemical composition within each ovum, we performed additional experiments on two *Rc* ova, in particular on the ova which produced the spectra (g) and (h) shown Fig. 2A. As shown in Fig.3, after three arbitrary repositioning of ovum (g) and ovum (h) we observe some variations of the linewidths as well as of the signal amplitudes. However, these experimental results clearly exclude that the origin of the observed diversity is only due to the non-homogeneous coil sensitivity combined with a non-uniform intracellular chemical composition. In other words, these experiments suggest that the heterogeneities in the absolute signal amplitudes as well as in the relative signal amplitudes (e.g. at about 3.8 ppm) cannot be attributed only to the positioning of the ovum but must be caused, at least partially, by its intrinsic properties. Further inspections of the experimental results via simple rescaling of the spectra (Figs. S3, S4, S5) confirm that both absolute and relative signal amplitudes are different in single ova, which are thus heterogeneous in the amount of intracellular compounds as well as in the chemical composition. Although our experimental conditions are very different from a well-controlled standard embryonic culture, it is not excluded that the observed heterogeneity is an observable consequence of developmental facts (see discussion in the Perspectives section).

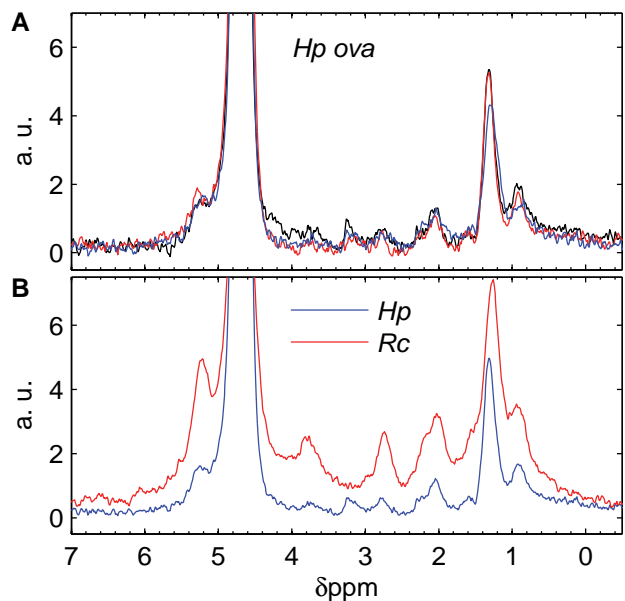

Fig. 4. NMR spectra of single *Heligmosomoides polygyrus bakeri* (*Hp*) ova and averaged spectra of *Richtersius coronifer* (*Rc*) and *Hp*. (A) Three *Hp* single ovum experiments in D$_2$O based gels. (B) Comparison between average spectra of five *Rc* ova (red) and three *Hp* ova (blue) in D$_2$O based gels.

Figure 4A shows $^1$H NMR spectra, resulting from 36 hours of averaging, obtained from three different single *Hp* ova placed in a 1.5% agarose D$_2$O gel. Contrarily to the case of *Rc* ova, in these measurements on *Hp* ova we do not observe a clear indication of heterogeneity. Due to the smaller volume (0.1 nl instead of 0.5 nl) not entirely compensated by a longer measuring time (36 hours instead of 12 hours), the lack of a clear evidence of heterogeneity in *Hp* ova might be a consequence of the worse signal-to-noise ratio and not an intrinsic property of the sample.

We conclude this section by comparing the NMR spectra of *Rc* and *Hp* ova. Figure 4B shows NMR spectra obtained from experiments on *Rc* and *Hp* ova in D$_2$O based gels. These spectra indicate that, at our level of sensitivity, the intracellular chemical composition of *Rc* and *Hp* ova are similar, with only small eventual differences (e.g. at 3.2 ppm) between the two species (see also Fig. S6).

**Perspectives**

The experiments reported in the previous section indicate the limitations of NMR spectroscopy for the study of subnanoliter single biological entities. In our specific case, the combination of a small number of spins (about 10 to 40 pmol of $^1$H nuclei for the 1.3 ppm signal) with a poor spectral resolution (about 50 Hz) determines a very long measuring time. Nevertheless, these experimental results suggest that, using miniaturized high sensitivity NMR probes, it is possible to point out differences in the intracellular chemical composition of single ova, showing the interest in performing experiments on single biological entities instead of ensembles. In our experimental conditions (essentially characterized by a field of 7 T, a microcoil having a sensitive volume of about 0.25 nl, and a spectral resolution of 50 Hz), an effective signal to noise ratio of three is achieved on about 300 pmol of $^1$H nuclei per single scan. Operation at higher fields would allow shorter experimental times. Moving from 7 T to 23.5 T (the highest field commercially available) should improve the spin sensitivity by a factor of six (see Appendix), allowing for 36 times faster measurements. In these conditions, we would achieve limits of detection on $^1$H nuclei in the order of 3 pmol in 10 minutes and 0.4 pmol in 10 hours.

In what follows we discuss the potential of our approach for intracellular chemistry studies of single mammalian zygotes. Studies of extracellular culture media using fluorometry and radio-labeled substrates reported that in sheep (23) and human (24) oocytes the uptake or production rates of metabolites such as lactate, pyruvate, and glucose can reach 100 pmol/oocyte/h and change radically along the natural development. It is worth to note that these results concern exchange rates measured in the extracellular medium and, hence, do not provide a direct quantification of the intracellular content and its time evolution. Spectrophotometry of intracellular extracts from ensemble of oocytes has shown that up to 30 pmol/oocyte of glutathione (GSH) are contained in oocytes of goat (25) and pig (26, 27) and can change in reaction to environment and developmental stage (28). Variations of few pmol/oocyte of GSH in time scales in the order of several hours have been reported in hamster (29) and rat (30) oocytes. In these studies the intracellular GSH content and its evolution (in connection with its important role in the embryonic development) is directly measured, but the ensemble measurements hide possible heterogeneities among single entities. These findings indicate that the sensitivity achievable with high sensitivity miniaturized inductive NMR probes should be sufficient for a non-invasive real-time intracellular monitoring of GSH in single mammalian zygotes. The application of NMR spectroscopy to the analysis of spent culture media was recently proposed to aid the selection of viable human embryos for in vitro fecundation purposes (31). The direct application of NMR on single embryos using miniaturized high sensitivity probes is probably advantageous for this aim. We suggest that systematic and extensive NMR studies on single cultured ova may provide new data that could shed light on cryptic processes involved in embryonic development (23-30) and provide new methodologies to estimate embryonic health (31).

Recently, many efforts were successfully dedicated to the microfabrication of devices for manipulation and isolation of living embryos (32). Both integrated circuits and microfluidics are suitable for arrays implementation, and their combination has been demonstrated in applications such as single cell magnetic manipulation (33) and flow cytometry (34). We believe that this combination can be extended to NMR applications and lead to the realization of arrays of high sensitivity NMR probes for simultaneous studies on large numbers of single biological entities.

## Materials and Methods

*Richtersius coronifer.* Eggs of *Rc* were extracted from a moss sample collected in Öland (Sweden) by washing the substrate, previously submerged in water for 30 min, on sieves under tap water and then individually picking up eggs with a glass pipette under a dissecting microscope. The eggs were shipped within 24 hours in sealed tubes with water and subsequently stored at -20 °C before use. All experiments were carried out within a week after tube opening. The tube was stored at 4 °C between separated experiments. The NMR experiments were performed in H$_2$O, M9, and D$_2$O. Prolonged exposure to a high concentration of D$_2$O seems to affect living organisms to different extents, from lethal to marginal (35, 36). In order to test the effects of D$_2$O exposure on *Rc* specimens, 16 eggs and 10 animals were submerged in D$_2$O (at 15°C) for 36 and 24 hours respectively and then transferred in H$_2$O. A control group of 16 eggs was kept in H$_2$O. The effects of the exposition to D$_2$O on the survival of the specimens were not negligible but definitely not systematically lethal: all the animals survived, while we observed a hatching of 84% in the control and of 63% in the eggs exposed after a time of observation of approximately 2 months.

*Heligmosomoides polygyrus bakeri.* Eggs of *Hp* were collected from faeces of infected mice. Faeces were first dissolved in water and then washed with saturated NaCl solution. Floating eggs were collected from the top layer of the solution and washed twice. Final centrifugation in water for 5 minutes at 2000 rpm sedimented clean eggs at the bottom of the tube. All experiments were carried out within two days after sample extraction. The tube was stored at 4 °C between separated experiments. The *Hp* ova regularly hatched in H$_2$O gels after a few hours, while no hatching was observed in D$_2$O gels. The use of D$_2$O allowed for 36 hours of averaging, a necessary condition to observe compounds other than H$_2$O in the 0.1 nl volume of the samples.

**NMR experimental details, data acquisition, and processing.** NMR experiments are performed in the 54 mm room temperature bore of a 7 T (300 MHz) superconducting magnet from Bruker. The electronic setup is identical to the one described in details in Ref. (20). All experiments are performed with repetition time of 2 s, $\pi/2$ pulse length of 2.5 μs, acquisition time of 400 ms, and sampling frequency of 650 kHz. The time domain data are post-processed applying an exponential filter with decay of 50 ms. The data shown in this work result from a total of 17 experiments involving three *Hp* ova and eight *Rc* ova, each one lasting for 12 to 36 hours. The alphabetic order in Fig. 2A corresponds to the chronologic order of the measurements.

**Chemicals.** H$_2$O (Sigma Aldrich, 270733). D$_2$O (Acros Organics, 166301000). Agarose (BioConcept, Standard Agarose type LE 7-01P02-R). Silicone glue (Momentive, RTV118). Polystyrene cup (Semadeni, 10 mm diameter, 5 mm height). The M9 buffer is prepared as in Ref. (37).

## Appendix
**Extrapolation of the sensitivity at higher fields.** The free induction decay (FID) NMR signal consists of several contributions at slightly different frequencies. Each contribution has an angular frequency $\omega_0 = \gamma_{eff} B_0$, where $\gamma_{eff}$ is the effective gyromagnetic ratio of the probed nuclei and depends on their chemical environment. In general, the signal contribution at each frequency is $s(t) = s_0 \cos(\omega_0 t) \exp(-t/T_2^*)$, where $s_0 = \gamma_{eff} \chi_0 B_0^2 V_s B_u / \mu_0$, $\chi_0$ is the static magnetic susceptibility of the probed nuclei, $\mu_0$ is the vacuum permeability, $V_s$ is the sample volume, $B_u$ is the magnetic field generated in the detection coil by a current of 1 A (i.e., the so called unitary field), $T_2^*$ is the effective decay time depending on the spin-spin relaxation time and the field inhomogeneity in the sample volume (38). The signal-to-noise ratio (*SNR*) for a single scan is maximized multiplying the time-domain signal by an exponential function whose decays matches $T_2^*$ (39). After this operation and for an acquisition time $T \gg T_2^*$, the *SNR* in the frequency domain is

$$SNR = \frac{s_0}{\bar{n}}\sqrt{\frac{T_2^*}{2}}\left(1-\exp\left(-\frac{2T}{T_2^*}\right)\right) \simeq \frac{s_0}{\bar{n}}\sqrt{\frac{T_2^*}{2}}$$

where $\bar{n}$ is the noise spectral density (39). Assuming that the measured linewidths are due to field distortions caused by susceptibility mismatches, the linewidth increases linearly with the static field (i.e., $T_2^* \propto 1/B_0$). Since, as shown above, the signal $s_0$ scales quadratically with the static field (i.e., $s_0 \propto B_0^2$), the *SNR* scales as $B_0^{3/2}$. Hence, if the field changes from 7 T to 23.5 T, we expect an *SNR* improvement of about six. In this extrapolation of the sensitivity at higher fields, we assumed that the noise spectral density does not depend on the operating field and that the same coil (i.e., same $B_u$) can be used. This is experimentally verified for our integrated detector up to 23.5 T (1 GHz). More generally, it is a good assumption for microcoils having a self-resonance frequency above the operating frequency and a coil metal thickness similar or smaller than the skin depth.

**Acknowledgments**

We thank Pierre Gönczy and Enrica Montinaro for discussions on the experimental setup. Ingemar Jönsson for providing the moss with *Rc*. Manuel Kulagin for helping in *Hp* ova extraction.

# Supporting Information

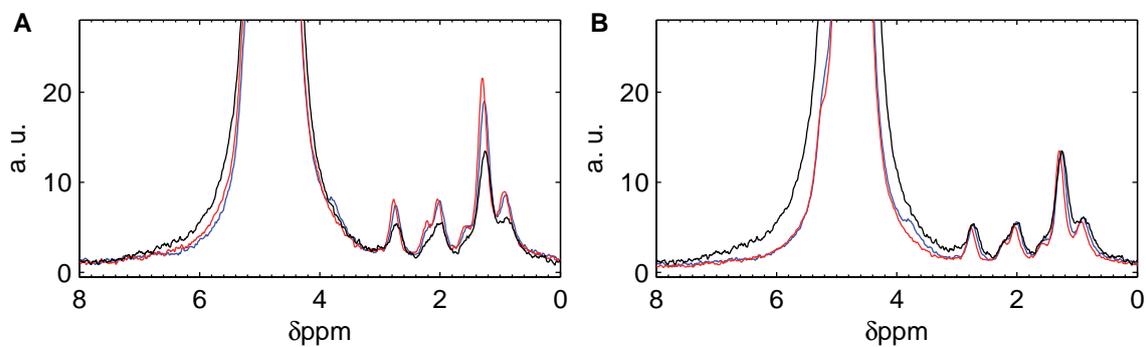

Fig. S1. Comparison between single *Rc* ova in H$_2$O based gel. Spectra of ova (a), (b) and (c) as indicated in Fig. 2A (main text). (A) Raw data (B) Spectra rescaled normalizing the peaks at 1.3 ppm.

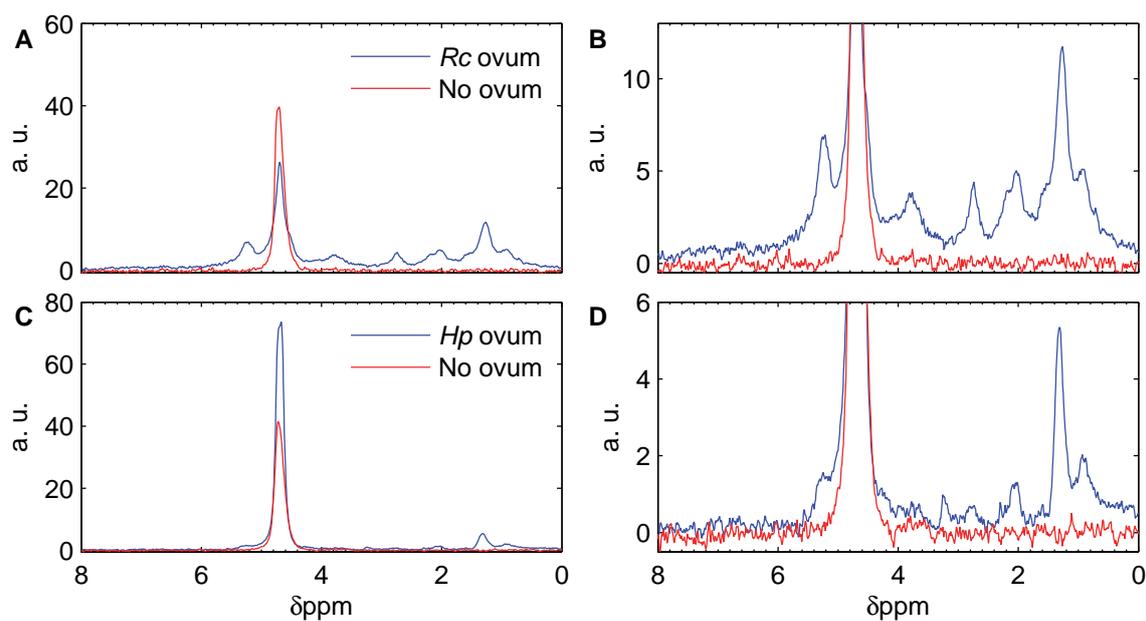

Fig. S2. Comparison between single ova experiments in D$_2$O based gel and D$_2$O based gel background. (A) Spectra of a single *Rc* ovum in D$_2$O based gel and of a D$_2$O based gel. (B) Detailed comparison. (C) Spectra of a single *Hp* ovum in D$_2$O based gel and of a D$_2$O based gel. (D) Detailed comparison.

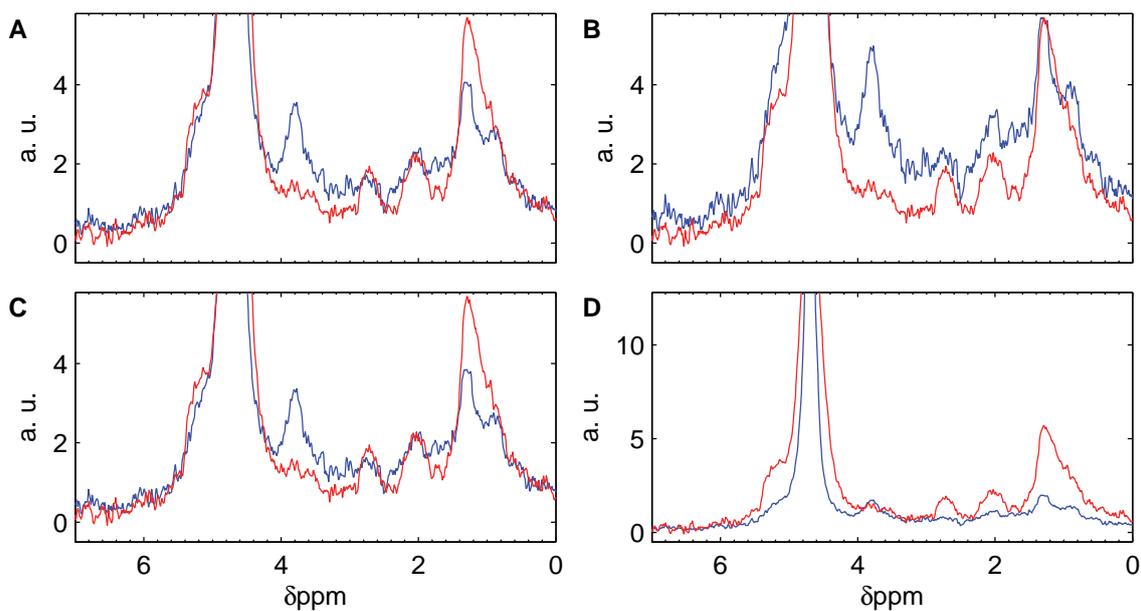

Fig. S3. Scaled spectra of single *Rc* ova. Spectra of ova (d) and (e) as indicated in Fig. 2A (main text). (A) Raw data. (B) Spectra rescaled normalizing the peaks at 1.3 ppm. (C) Spectra rescaled normalizing the peaks at 2 ppm. (D) Spectra rescaled normalizing the background peaks at 4.7 ppm.

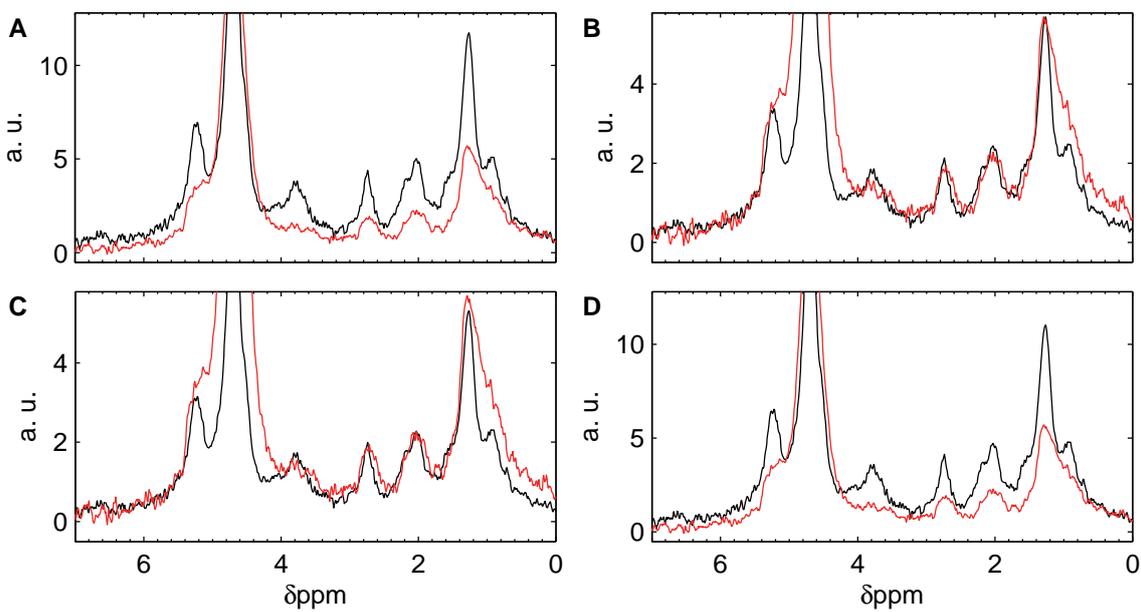

Fig. S4. Scaled spectra of single *Rc* ova. Spectra of ova (e) and (f) as indicated in Fig. 2A (main text). (A) Raw data. (B) Spectra rescaled normalizing the peaks at 1.3 ppm. (C) Spectra rescaled normalizing the peaks at 2 ppm. (D) Spectra rescaled normalizing the background peaks at 4.7 ppm.

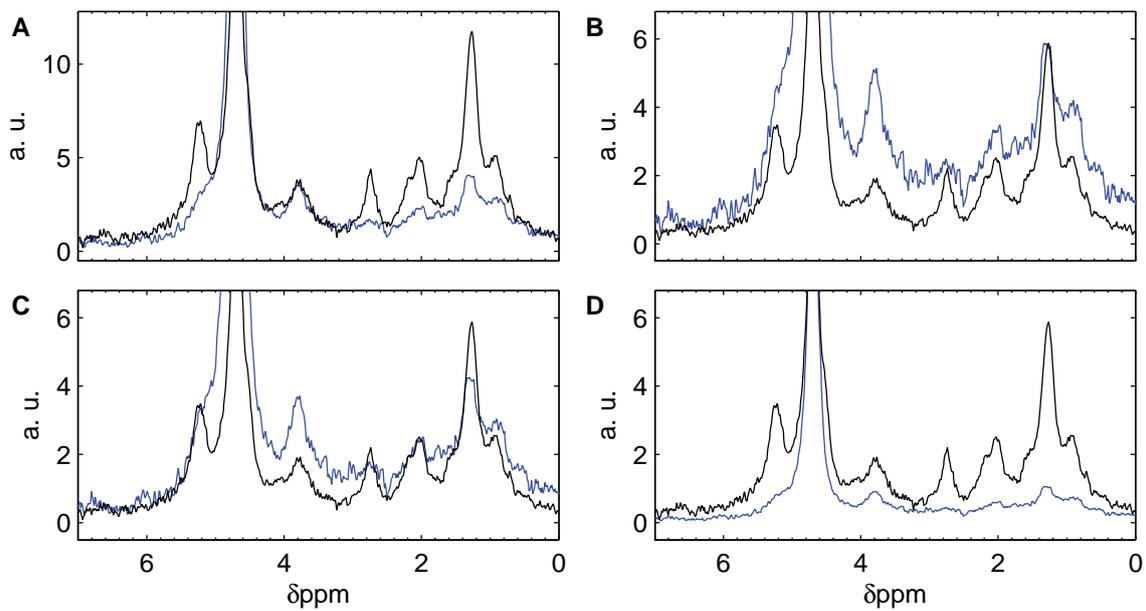

Fig. S5. Scaled spectra of single *Rc* ova. Spectra of ova (d) and (f) as indicated in Fig. 2A (main text). (A) Raw data. (B) Spectra rescaled normalizing the peaks at 1.3 ppm. (C) Spectra rescaled normalizing the peaks at 2 ppm. (D) Spectra rescaled normalizing the background peaks at 4.7 ppm.

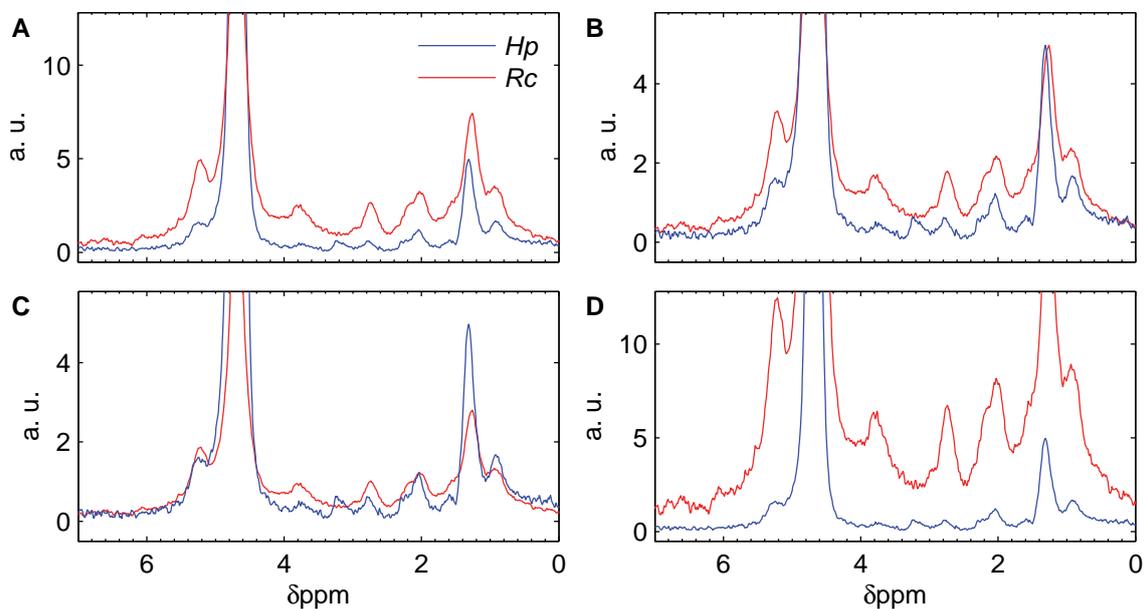

Fig. S6. Scaling of averaged *Hp* (blue) and *Rc* (red) spectra from experiments in $D_2O$ based gels. Averaged spectra of *Rc* and *Hp* ova as indicated in Fig. 3B (main text). (A) Raw data. (B) Spectra rescaled normalizing the peaks at 1.3 ppm. (C) Spectra rescaled normalizing the peaks at 2 ppm. (D) Spectra rescaled normalizing the background peaks at 4.7 ppm.